# Investigation of mechanical properties of cryogenically treated music wire

A Heptonstall[+], M Waller and N A Robertson*

LIGO Laboratory, California Institute of Technology, Pasadena, CA 91125, USA

[+]Corresponding author email: heptonstall_a@ligo.caltech.edu
* Also SUPA, School of Physics and Astronomy, University of Glasgow, Glasgow, G12 8QQ, Scotland

It has been reported that treating music wire (high carbon steel wire) by cooling to cryogenic temperatures can enhance its mechanical properties with particular reference to those properties important for musical performance. We use such wire for suspending many of the optics in Advanced LIGO, the upgrade to LIGO - the Laser Interferometric Gravitational-Wave Observatory. Two properties that particularly interest us are mechanical loss and breaking strength. A decrease in mechanical loss would directly reduce the thermal noise associated with the suspension, thus enhancing the noise performance of mirror suspensions within the detector. An increase in strength could allow thinner wire to be safely used, which would enhance the dilution factor of the suspension, again leading to lower suspension thermal noise. In this article we describe the results of an investigation into some of the mechanical properties of music wire, comparing untreated wire with the same wire which has been cryogenically treated. For the samples we studied we conclude that there is no significant difference in the properties of interest for application in gravitational wave detectors.

## I. INTRODUCTION

Suspension thermal noise [1-4] is one of the main sources of noise which could limit the low frequency performance of gravitational wave detectors such as Advanced LIGO, VIRGO, Kagra and GEO600 [5-8]. These detectors use interferometry between suspended masses, called test masses, to look for strains in space produced by gravitational waves from astrophysical sources. To reduce the level of suspension thermal noise in these detectors the test masses are suspended by materials with low mechanical loss. Advanced LIGO, VIRGO and GEO600 use silica fibers, and Kagra will use sapphire fibers, both of which materials have losses approximately three orders of magnitude better than steel music wire [9-11]. However auxiliary mirrors, whose requirements on noise performance are less than for the test masses, are typically suspended using metal wire.  In Advanced LIGO these mirrors include the beamsplitter [12], the mirrors which form the modecleaner cavity which improves the beam quality of the input laser light, and the mirrors used in the two recycling cavities which enhance the power level and signal level in the detector.  All of these suspensions contribute at some level to the overall noise performance of the detector and in some cases the expected thermal noise performance is close to or even at the requirement level for those suspensions. This can be seen for example in the beamsplitter suspension design, see figure 1 of Robertson and Barton [12]. In this figure the expected displacement noise due to suspension thermal noise is shown, estimated assuming a wire loss of $2 \times 10^{-4}$, based on Cagnoli et al. [11]. It can be seen to lie essentially at the requirement level in the 20 to 40 Hz region. Any improvement in this value of wire loss would take the noise below the requirement.

We use music wire rather than another metal for these suspensions due to its intrinsic high tensile strength (up to 3 GPa depending on diameter) as well as its good handling properties [13]. Any improvement in the mechanical properties and hence thermal noise performance of this wire could be beneficial.

Chen et al. [14] report on investigations of samples of guitar strings G1 to G6, with diameters ranging from 1.35 mm to 0.30 mm, in which improved characteristics of music wire were seen after cryogenic treatment. The treatment involved cooling slowly to -184 °C, leaving the strings for 30 hours at that temperature and slowly increasing again (see their paper for details). They report an increase in Young's modulus (up to 80% increase) and tensile strength (up to 8% increase), and a reduction in creep rate (typically more than 10%) and hardness. They note that there is a relationship between creep due to grain boundary realignment and the Köster effect [15,16], a brightening of string sound attributed to a gradual decay of internal friction of drawn material under tension over time. Thus they conclude that a decrease in creep will correspond to a decrease in internal friction (mechanic loss). However there are no direct measurements of mechanical loss in this paper or in the referenced thesis [17].

They note that there is no clear-cut understanding of the mechanism(s) which lead to the improvements. Possible contributing factors are dispersion strengthening phenomena associated with the cryogenic treatment as well as stress relief accompanying precipitate formation.

The improvements for all properties observed by Chen et al. are not dramatic. However any decrease in suspension thermal noise could be useful for our application, and hence we have carried out further investigations, in particular including measurements of mechanical loss. We note that the guitar strings investigated by Chen et al. consisted of a steel core with bronze winding (G1 to G4) or steel wire with a thin layer of tin plating (G5 and G6). We are interested in the properties of steel music wire with no winding or coating. Music wire normally comes with a phosphate coating to reduce rusting. For Advanced LIGO we procured uncoated music wire as a custom item. We have strict requirements on which materials can be used in the ultrahigh vacuum system for Advanced LIGO and a phosphate coating is not acceptable. Thus we may expect to see different behavior than previously reported.

The cryogenically treated wires tested here were treated by CoolTech [18] in Austria. The wire is cycled to cryogenic temperatures several times without coming in contact with the coolant, reaching a low temperature of -180 °C. The process takes 15hrs, during which time the temperature cycle is computer controlled to keep the ramp rate between 1-2 °C/min and to allow reproducible results. The exact cooling cycle used is proprietary. We note that the lowest temperature achieved is similar to Chen et al. However the cycling of temperature is different, as is the overall time of the process. The low ramp rate of temperature is used to keep thermal gradients low during the cycling.

In section II we introduce the key theory and equations for suspension thermal noise to put our investigations into perspective. In Section III we describe our wire samples and cover the breaking strength experiments and results. Section IV presents the mechanical loss experimental method and results. In section V we discuss our results and present our conclusions.

## II.     SUSPENSION THERMAL NOISE

Thermal noise in a mirror suspension can be characterized by the fluctuation dissipation theorem [1,2]

$$S_F(\omega) = 4k_b T \mathbb{R}[Z(\omega)], \qquad (1)$$

where $S_F(\omega)$ is the power spectral density of the thermal driving force, $k_B$ is the Boltzmann constant, $T$ is the temperature, $\omega$ is the angular frequency and $\mathbb{R}[Z(\omega)]$ is the real part of the admittance.

At angular frequencies, $\omega$, well above resonance, the spectral density of displacement noise, $S_X(\omega)$, arising from the suspension wire can be written as

$$S_X(\omega) = \frac{4k_B T \omega_0^2}{m\omega^5} \phi(\omega), \qquad (2)$$

where $m$ is the suspended mass, $\omega_0$ is the angular resonant frequency and $\phi(\omega)$ is the loss angle.

The suspension wire loss angle is given by

$$\phi(\omega) = D(\phi_t(\omega) + \phi_m), \qquad (3)$$

where $\phi_t(\omega)$ is the thermoelastic loss angle [19] arising from thermal expansion, and $\phi_m$ is the material loss angle arising from internal friction. $D$ is the dilution factor [20], which is a reduction in loss by an amount equaling the ratio of energy stored in lifting the mass to the energy stored in bending the wire.

For a wire that is not under load the thermoelastic loss may be calculated from [21]

$$\phi_t = \Delta \frac{f/f_r}{1+(f/f_r)^2}, \qquad (4)$$

where the magnitude of damping is given by

$$\Delta= \left(\frac{\alpha Y T}{\rho C}\right). \qquad (5)$$

$\alpha$ is the thermal expansion coefficient, $C$ is the specific heat capacity, $Y$ is the Young's modulus, $\rho$ is the density and $f_r$ is the characteristic frequency for heat transfer across a wire of diameter $d$, given by

$$f_r = \frac{2.16\kappa}{\rho C d^2}, \qquad (6)$$

where $\kappa$ is the thermal conductivity.  For a wire that is under tension an additional term is added that arises from the temperature dependence of the Young's modulus.  For a freely hanging wire this term disappears due to there being no static extension of a wire under no load.  Also the dilution factor term becomes 1, giving

$$\phi_{unloaded}(\omega) = \phi_t(\omega) + \phi_m \qquad (7)$$

By directly measuring the loss angle of the music wire the suspension thermal noise can be calculated using Eqns. 1 to 7.  The loss angle can be measured using the ringdown technique described in section IV(A).

## III. STRENGTH MEASUREMENTS

### A. Experimental Method

Destructive strength testing was performed on wires supplied by California Fine Wire Company with a range of diameters from 20 µm to 1100 µm.  In each case the wire length was kept to 0.5 m.  The testing was performed using a commercial Instron strength testing machine, allowing recording of force vs extension as a function of time.  The wire was clamped at both ends using self-tightening wedge style clamps.  In practice the machine works by applying a uniform rate of extension and measuring the force using a load cell attached to one of the clamps.  In total 43 untreated wires were tested, and 7 cryogenically treated wires. Figure 1 shows a typical extension vs force measurement.  Initially the extension is a linear function of applied force.  A mean value of Young's modulus of $2.12 \pm 0.01 \times 10^{11}$ Pa was found for the untreated wire, and $2.00 \pm 0.12 \times 10^{11}$ for the cryogenically treated wire.  These are both consistent with previous measurements made on untreated wire.  After a certain force is reached the wire begins to draw down in diameter, as it begins to deform plastically.  This is seen as the non-linear region in the figure.  A small non-linear region is seen around 210 N in each measurement and was found to be due to a small amount of play in the attachment of the lower clamp.  Eventually sufficient force is applied that the breaking strength is reached and the force drops to zero.

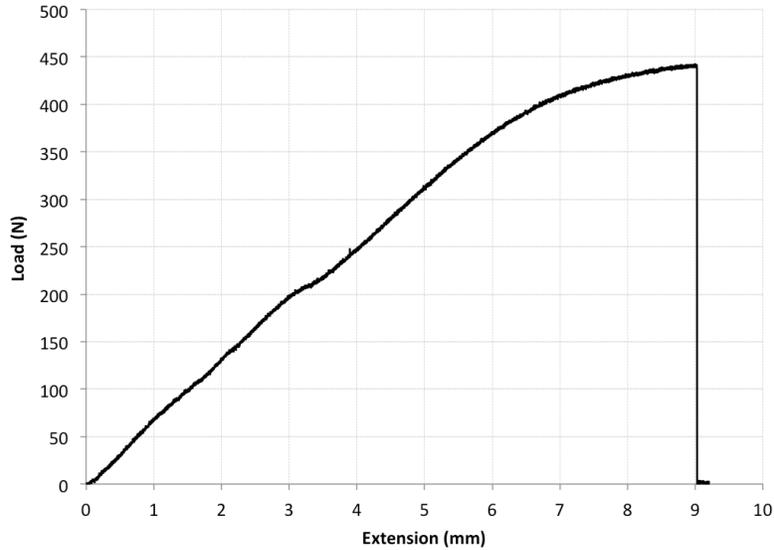

Figure 1: Plot of destructive strength testing of a 450µm wire. The linear region can clearly be seen up until an extension of approximately 6.5mm, followed by plastic deformation, and the break at just below 9mm extension.

### B.    Strength Measurement Results

Table 1 lists the diameters of the wires tested, along with average breaking strengths. Figure 2 shows the results of the destructive strength testing along with data from the ASTM standard for maximum and minimum expected strengths for high carbon steel wire [13]. The ASTM standard gives two possible clamping configurations, one of which, the self-tightening wedge clamps, we used for our tests. The ASTM standard also calls for samples that broke in the clamp to be discarded. We have denoted any wire that broke in or close to the clamps as an asterisk in Figure 2. In practice we see no pattern that these wires had lower strength, and the data from the strength tests shows plastic deformation similar to the other wires before the ultimate strength was reached. It is interesting to note that some of the highest strengths measured were on wires that broke in or close to the clamp.

| Diameter (µm) | Number of samples | Treated / Untreated | Average breaking stress (GPa) |
|---|---|---|---|
| 1100 | 7 | Untreated | 2.38 |
| 711 | 2 | Untreated | 2.52 |
| 635 | 2 | Treated | 2.17 |
| 635 | 1 | Untreated | 2.16 |
| 609 | 2 | Untreated | 2.13 |
| 457 | 2 | Treated | 2.52 |
| 457 | 2 | Untreated | 2.58 |
| 450 | 16 | Untreated | 2.82 |
| 355 | 2 | Untreated | 2.60 |
| 340 | 2 | Untreated | 2.36 |
| 269 | 2 | Untreated | 2.07 |
| 248 | 2 | Untreated | 2.13 |
| 243 | 2 | Untreated | 2.17 |
| 203 | 1 | Untreated | 2.23 |
| 197 | 3 | Treated | 2.49 |
| 197 | 2 | Untreated | 2.01 |

Table 1: Sample data for destructively strength tested wires.

We find no evidence that the cryogenically treated wire differs in absolute strength from the non-treated wire. We note also that compared to the ASTM standard for maximum and minimum breaking stresses, many of our data points were outside of these values, particularly for thinner wires. In practice we are also interested to know the point at which the material begins to exhibit plasticity, since suspended optics must remain in the linear elastic region. Since the machine applies a uniform rate of extension to the wire, plasticity is seen as a reduction in the measured force for a given extension of the wire. To find this point, a fit was made to linear region of the plots of extension against force. We define a point at which the material has deviated from this linear behavior by finding the extension at which the applied force has deviated three standard deviations from the value given by the linear fit. On average the linear limit is reached at 85% of the breaking stress, and this was highly reproducible between samples. Again we find that there is no significant difference between cryogenically treated and untreated wire.

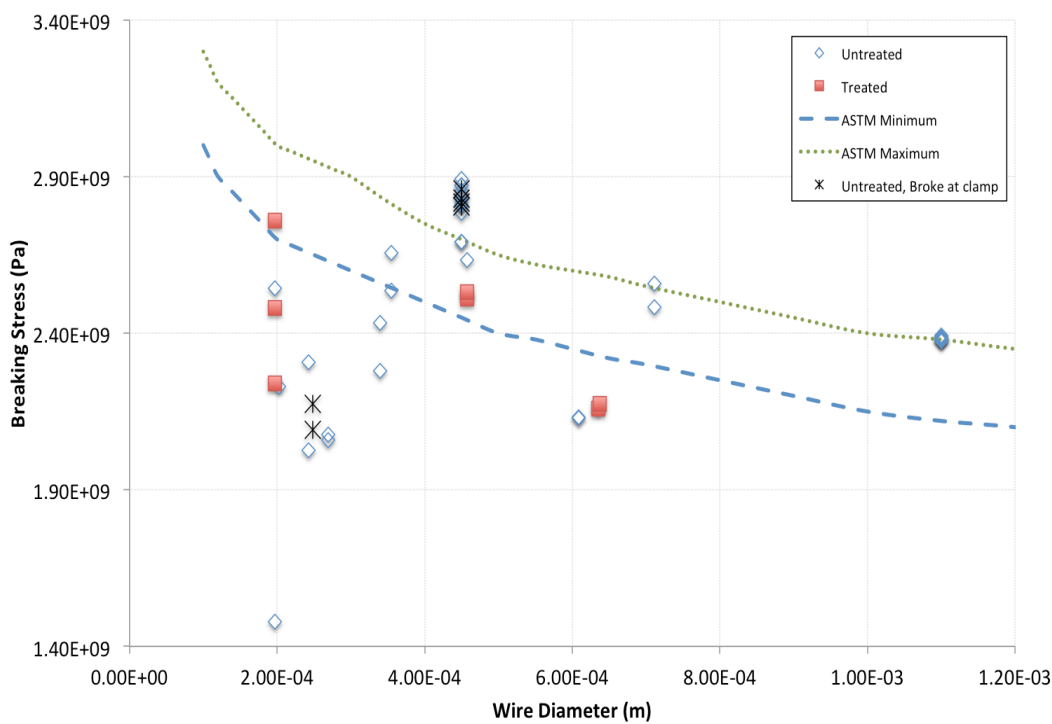

Figure 2: Measured breaking strengths of high carbon steel wire. Untreated wire is shown as blue diamonds. Cryogenically treated wire is shown as orange squares. Untreated wire that broke in or at the clamp is shown as asterisks. ASTM standard data [13] is also shown for minimum (blue dashed line) and maximum (green dotted line) strengths.

## IV. MECHANICAL LOSS MEASUREMENTS

### A. Experimental Method

Mechanical loss measurements were performed on each of four wire samples. A ringdown technique was used in which the wire is clamped at one end using a stainless-steel, Advanced-LIGO design of wire clamp which features a v-shaped groove to constrain the direction of the clamped end. The clamp in turn was bolted to a stiff structure and suspended in a vacuum chamber. The resonant modes of the wire were excited using an electrostatic force and were allowed to decay, with the amplitude of the decaying sinusoid being measured using a HeNe laser and split photodiode sensor. The time constant for the decay was then found by fitting to the ringdown, and the loss angle calculated from this. The wires were chosen such that two were cryogenically treated and two were untreated. Diameters of wire were chosen to reflect those used in aLIGO suspensions, and the diameters were measured along the lengths of the wire using a screw gauge micrometer. Table 2 below shows wire dimensions, and whether the wire was already straightened or was hand straightened for this test.

| Wire number | Diameter | Treatment | Length (mm) | Pre-straightened? |
|---|---|---|---|---|
| 1 | 465µm | untreated | 498 | straight |
| 2 | 457µm | cryo treated | 497 | straightened |
| 3 | 197µm | untreated | 473 | straightened |
| 4 | 197µm | cryo treated | 481 | straightened |

**Table 2: Dimensions of tested wires**

The apparatus used for the ringdown measurement is shown in Figure 3.

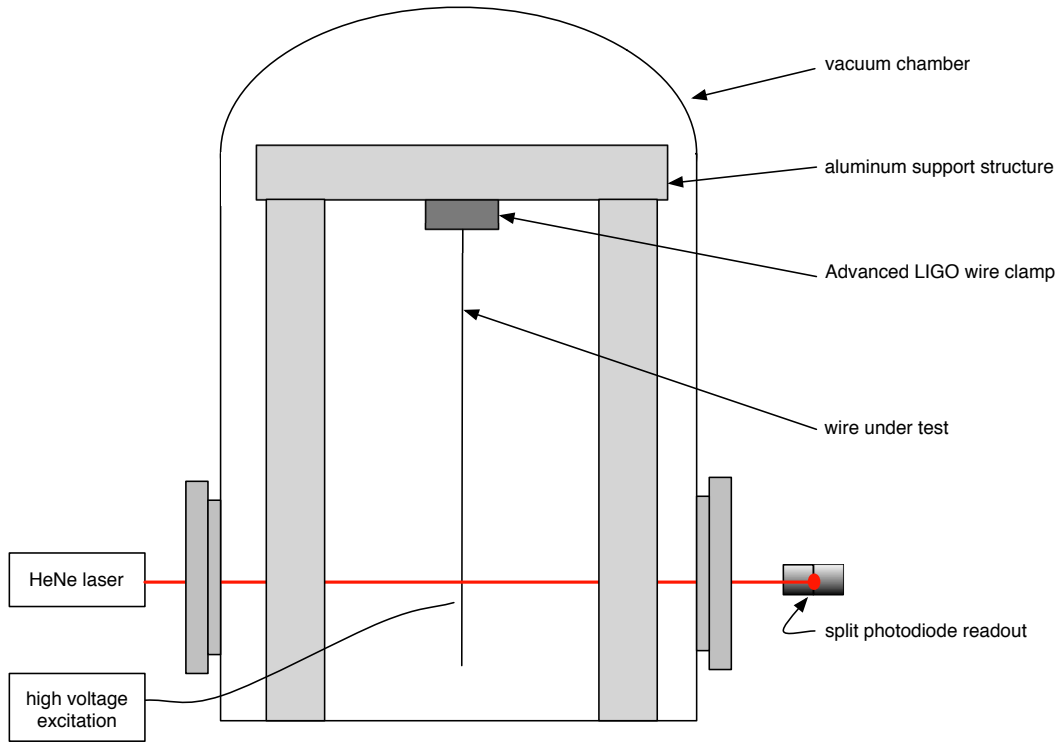

**Figure 3: Apparatus for measurement of time constants of mechanical mode ringdowns.**

We note that three of the samples measured were delivered wound onto spools and had to be straightened before testing. The wires were straightened by hand by bending until they deformed plastically. The primary reason for straightening the wire was to allow it to be hung for Q measurement. After hand straightening the wires showed residual curvature over shorter regions of approximately 1 inch, however the radius from the wire spool was on average removed such that the top and bottom of the wire were inline when clamped to within a few inches. It can be assumed that this process will induce some form of work hardening of the wire. We see no evidence in the frequencies of modes that this has changed the Young's modulus significantly. It might be expected that dislocations induced by work hardening would increase the mechanical loss, though we see no evidence that the straightening process used here has altered the losses in the untreated wires when compared to the already straight wire. The work hardening experienced during straightening is likely to have been small, since no reduction in diameter was found after straightening, this being a typical indicator of the hardening. We note that in practice any wire hung optics would be suspended on wire taken from a spool but not straightened since the wire is never subjected to forces above the plastic limit.

### B. Mechanical Loss Measurement Results

For each wire a range of resonant modes were found and these were plotted against the predicted mode frequencies, using the diameters in Table 2 and material values from Table 3. This ensures the measured modes are the cantilever modes. Also this gives a check on the Young's modulus and density values for the material, which we find do not change significantly after cryogenic treatment. An example of this for the 457μm cryogenically treated wire is shown in Figure 4.

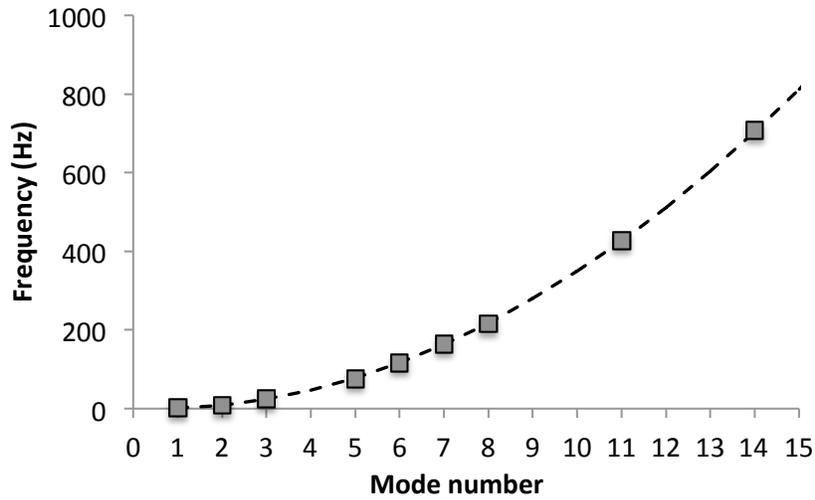

Figure 4: Resonant modes of a 457μm wire shown as squares. The theoretical values, based on material values given in Table 3, are plotted as a dashed line.

| $Y(N/m^2)$ | $\alpha(K^{-1})$ | $\rho(kg/m^3)$ | $k(W/mK)$ | $C(J/K)$ | $\phi$ |
|---|---|---|---|---|---|
| $2.12\times10^{11}$ | $1.1\times10^{-5}$ | 7800 | 49 | 486 | $2\times10^{-4}$ |

Table 3: Material parameters for music wire used to predict thermal noise of wire suspended optics in Advanced LIGO

Figure 5 shows data for the measured loss values for wire number 4. The errors plotted are the standard error calculated from three individual ringdown measurements for each mode. The curve in this plot shows the predicted losses based on material values from the literature, and loss values currently used to predict thermal noise performance of metal wire suspended stages in Advanced LIGO, given in Table 3.

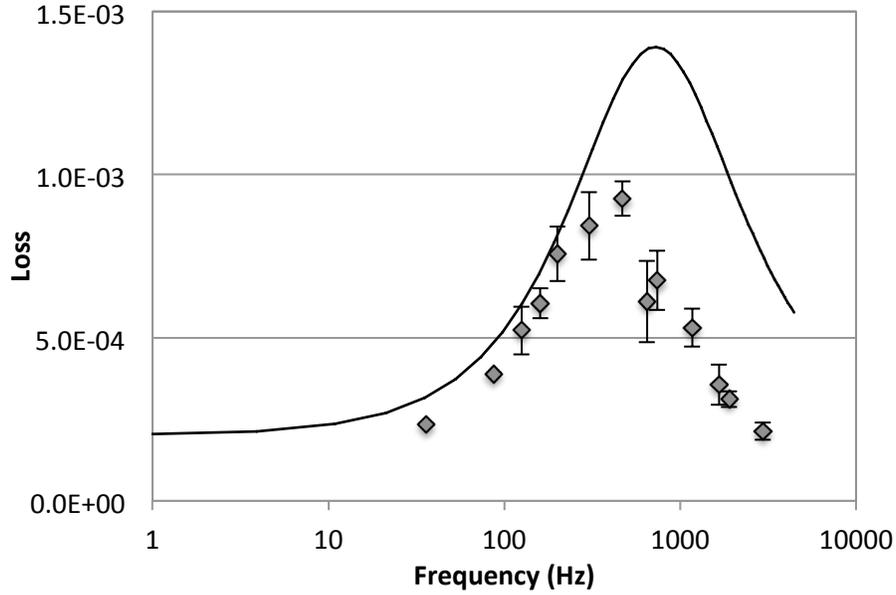

Figure 5: Measured losses for Wire 4 are shown as plotted points. The plotted errors are the standard error from separate measurements of each mode. The line shows the predicted loss based on material and loss values used to predict losses in Advanced LIGO, given in Table 3.

From the plots for each wire it was apparent that the total loss predicted was too high. It was also seen that for both thinner fibers the predicted peak thermoelastic frequency was too high. For the two thicker fibers the predicted peak thermoelastic frequency was closer to the data, though slightly too high. A linear regression fit was then performed allowing the material loss value to vary for each wire. The thermoelastic relaxation strength coefficient Δ and peak frequency $f_r$ were allowed to vary for the two different diameters of wire. A similar fit is seen in Cagnoli et al. [11]. While we know the Young's modulus and density values are correct, and that the dimensions of the wire are well known, our fit does not allow us to distinguish between $\kappa$, $C$ and $\alpha$. We allow the material parameters to be fixed for the two similar diameters of wire based on observation of the experimental data.

Figure 6 and Figure 7 show the plots of measured loss for each of the four wires along with the fitted lines. The 79 Hz mode of the 465 μm untreated wire and the 470 Hz mode of the 197 μm untreated were removed from their respective fits as they exhibit excess loss. The parameters found from the fits are given in Table 4.

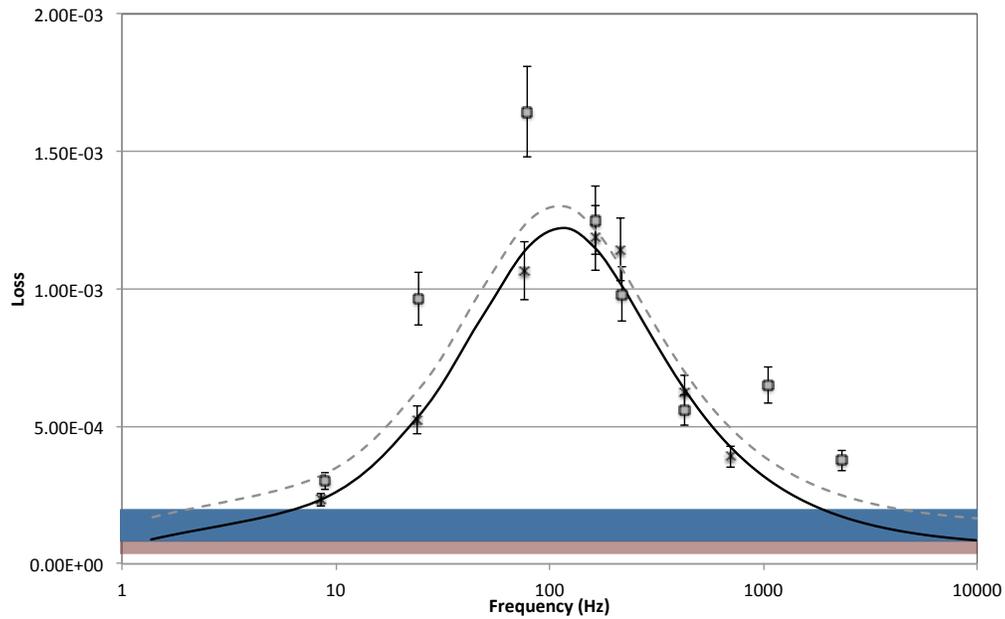

Figure 6: Loss measurements for 465μm untreated (square points) and 457μm cryogenically treated (asterisks). A fit to Eq. 6 is shown as a dashed line for the untreated wire and a solid line for the cryogenically treated wire. Shaded areas represent range of values of material loss possible for the two wires based on error analysis.

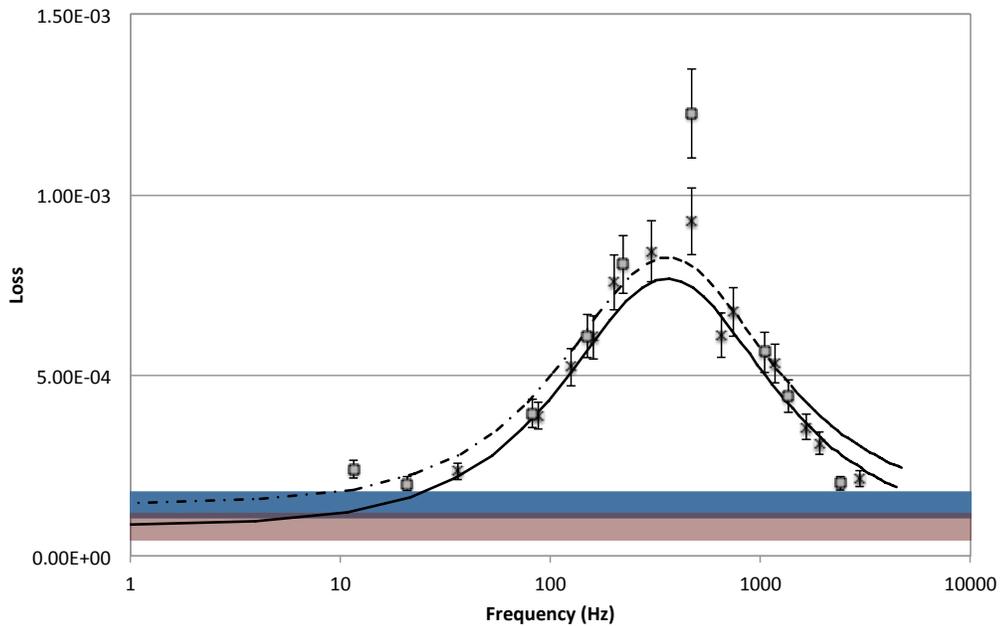

Figure 7: Loss measurements for 197μm untreated (square points) and 197μm cryogenically treated (asterisks). A fit to Eq. 6 is shown as a dash-dot line for the untreated wire and a solid line for the cryogenically treated wire. Shaded areas represent range of values of material loss possible for the two wires based on error analysis.

| Wire | Treatment | Mechanical Loss | Δ fit | $f_r$ fit(Hz) | Predicted $f_r$(Hz) | Predicted Δ |
|---|---|---|---|---|---|---|
| 1 | untreated | $1.4 \pm 0.6 \times 10^{-4}$ | $2.3 \pm 0.1 \times 10^{-3}$ | $109 \pm 9$ | 129 | $2.0 \times 10^{-3}$ |
| 2 | cryo-treated | $0.6 \pm 0.2 \times 10^{-4}$ | $2.3 \pm 0.1 \times 10^{-3}$ | $113 \pm 9$ | 134 | $2.0 \times 10^{-3}$ |
| 3 | untreated | $1.4 \pm 0.4 \times 10^{-4}$ | $1.4 \pm 0.2 \times 10^{-3}$ | $359 \pm 54$ | 719 | $2.0 \times 10^{-3}$ |
| 4 | cryo-treated | $0.8 \pm 0.4 \times 10^{-4}$ | $1.4 \pm 0.2 \times 10^{-3}$ | $359 \pm 54$ | 719 | $2.0 \times 10^{-3}$ |

Table 4: Values from fit to mechanical loss measurements

There is clear evidence both from these results and also from Cagnoli et al. [11] that material values for the thin wires measured here are different from those quoted in the literature for bulk pieces. It would be possible, though outwith the scope of this work, to directly measure thermal conductivity, specific heat capacity and thermal expansion coefficient such that these could be excluded from the fit. This would be expected to reduce the error in the fitted mechanical loss values. For both diameters of wire a decrease in fitted loss value is seen for the cryogenically treated samples, though the results do agree within the limits of experimental error. We therefore conclude there is no evidence of a significant reduction in material loss through cryogenic treatment. Comparing the fitted material loss values for untreated wires with those from [11], $\phi_{mat} = 1.9 \pm 0.1 \times 10^{-4}$, we find a slightly lower average value for wire of similar diameter at $\phi_{mat} = 1.4 \pm 0.4 \times 10^{-4}$, though we note that taking experimental error into account the results do agree. Assuming no effect from cryogenic treatment the mean material loss value for high carbon steel wire measured here is $1.1 \pm 0.2 \times 10^{-4}$. We note that the shift in magnitude and peak frequency of thermoelastic damping for thinner wires is larger than seen in Cagnoli et al. [11], however while the effect is more marked the shift in both cases is in the same direction. The thicker wire measured shows fit values closer to the predicted frequency and magnitude.

## V. CONCLUSIONS

The mechanical losses in cryogenically treated high carbon steel music wire were measured and compared to standard untreated music wire. Within the limits of experimental errors no significant reduction in internal friction was seen. However there was evidence of significant differences in the material properties of this thin wire, both treated and untreated, compared to bulk values taken from the literature. Destructive strength testing of cryogenically treated wire showed no evidence of changes to the ultimate strength. We conclude that cryogenically treating music wire does not reduce the internal friction or increase strength at a level that would yield improved thermal noise performance for the aLIGO mirror suspensions.

## ACKNOWLEDGEMENTS
The authors would like to thank Andrew Hoff for his assistance in using the Instron strength testing machine in the Keck Laboratories at Caltech, and Ben Abbott for his assistance in straightening samples of wire and his general advice on metallurgy. We would like to thank Jeff Lewis for his assistance in procuring the cryogenically treated wire samples. We would like to thank Peter Saulson and James Hough for their helpful discussions on aspects of this work.  We would also like to thank our colleagues in the LIGO Scientific Collaboration for their interest in this work.  We are grateful for the financial support provided by the National Science Foundation.   LIGO was constructed by the California Institute of Technology and Massachusetts Institute of Technology with funding from the National Science Foundation, and operates under cooperative agreement PHY-0757058. Advanced LIGO was built under award PHY-0823459.

REFERENCES

1. H. B. Callen, and T. A. Welton, "Irreversibility and Generalized Noise", Phys. Rev. **83**, 34-40 (1951)

2. H. B. Callen, and R. F. Greene, "On a Theorem of Irreversible Thermodynamics", Phys. Rev. **86**, 702-710 (1952)

3. P. R. Saulson, "Thermal noise in mechanical experiments", Phys. Rev. D **42**, 2437-2445 (1990)

4. S. Rowan, J. Hough, and D. R. M. Crooks, "Thermal noise and material issues for gravitational wave detectors", Phys. Lett. A **347,** 25-32 (2005)

5. J. Aasi, et al. (The LIGO Scientific Collaboration)  "Advanced LIGO",  Class. Quantum Grav. **32**, 074001 (2015)

6. F. Acernese, et al. (The Virgo Collaboration) "Advanced Virgo: a second-generation interferometric gravitational wave detector",  Class. Quantum Grav. 32 024001 (2015)

7. K. Somiya, "Detector configuration of KAGRA–the Japanese cryogenic gravitational-wave detector", Classical and Quantum Gravity **29**, 124007, (2012).

8. C. Affeldt, , K. Danzmann, K. L. Dooley, H. Grote., M. Hewitson, S. Hild, J. Hough, J. Leong, H. Lück, M. Prijatelj, S. Rowan, A. Rüdiger, R. Schilling, R. Schnabel, E. Schreiber, B. Sorazu, K. A. Strain, H. Vahlbruch, B. Willke, W. Winkler and H. Witte, "Advanced techniques in GEO 600", Class. Quantum Grav. **31**, 224002 (2014)

9. A. Heptonstall, M. Barton, C. Cantley, A. Cumming, G. Cagnoli, J. Hough, R. Jones, R. Kumar, I. Martin, S. Rowan, C. Torrie and S. Zech, "Investigation of mechanical dissipation in $CO_2$ laser-drawn fused silica fibers and welds", Class. Quantum Grav. **27**, 035013 (2010)

10. T. Uchiyama, T. Tomaru, D. Tatsumi, S. Miyoki, M. Ohashi, K. Kuroda, T. Suzuki, A. Yamamoto and T. Shintomi, "Mechanical quality factor of a sapphire fiber at cryogenic temperatures", Phys.Lett. **A273**, 310-315 (2000)

11. G. Cagnoli, L. Gammaitoni, J. Kovalik, F. Marchesoni and M. Punturo, "Low-frequency internal friction in clamped-free thin wires, Phys Lett A **255,** 230-235 (1999)

12. N. A. Robertson, and M. Barton "Conceptual Design of Beamsplitter Suspension for Advanced LIGO", 2009, https://dcc.ligo.org/LIGO-T040027/public

13. ASTM International, "Standard Specification for Steel Wire, Music Spring Quality", A 228/A/228M – 07 (2007)

14. J. M. Chen, K. H. W. Seah, and C. H. Chew, "Mechanical Characterization of Cryogenically Treated Music Wire" Journal of ASTM International, **3**, Issue 4 (April 2006)

15. A. S. Nowick, Internal Friction and dynamic modulus of cold-worked metals, Journal of Applied Physics, **25**, 1129-1134 (1954),

16. M. Goodway, Metals of Music, Materials Characterization, **29**, 177-184 (1992)

17. J. M. Chen, "Cryogenic Treatment of Music Wire" Master of Engineering Thesis, Department of Mechanical Engineering, National University of Singapore, (2004)

18. CoolTech, Blockau 64a, A-6642 Stanzach, Austria, http://www.cooltech.at/en/

19. A. S. Nowick and B. S. Berry, Anelastic Relaxations in Crystalline Solids, London Academic (1972)

20. G. Cagnoli, J. Hough, D. DeBra, M. M. Fejer, E. Gustafson, S. Rowan, V. Mitrofanov, "Damping Dilution Factor For a Pendulum in an Interferometric Gravitational Wave Detector", Phys. Lett. A **272**, 39-45 (2000)

21. G. Cagnoli and P. A. Willems, "Effects of Nonlinear Thermoelastic Damping in Highly Stressed Fibers", Phys. Rev. B, **65** 174111 (2002)